\pdfoutput=1
\documentclass{aa}  

\usepackage{graphicx}
\usepackage{txfonts}
\usepackage{natbib,twoopt}
\usepackage[breaklinks=true]{hyperref}
\bibpunct{(}{)}{;}{a}{}{,}             
\makeatletter
  \newcommandtwoopt{\citeads}[3][][]{\href{http://adsabs.harvard.edu/abs/#3}%
    {\def\hyper@linkstart##1##2{}%
     \let\hyper@linkend\@empty\citealp[#1][#2]{#3}}}
  \newcommandtwoopt{\citepads}[3][][]{\href{http://adsabs.harvard.edu/abs/#3}%
    {\def\hyper@linkstart##1##2{}%
     \let\hyper@linkend\@empty\citep[#1][#2]{#3}}}
  \newcommandtwoopt{\citetads}[3][][]{\href{http://adsabs.harvard.edu/abs/#3}%
    {\def\hyper@linkstart##1##2{}%
     \let\hyper@linkend\@empty\citet[#1][#2]{#3}}}
  \newcommandtwoopt{\citeyearads}[3][][]%
    {\href{http://adsabs.harvard.edu/abs/#3}
    {\def\hyper@linkstart##1##2{}%
     \let\hyper@linkend\@empty\citeyear[#1][#2]{#3}}}
\makeatother

\begin{document} 

\titlerunning{Testing the Existence of Event Horizons against Rotating Reflecting Surfaces}
\authorrunning{Joost de Kleuver et al.}

\title{Testing the Existence of Event Horizons against\\ Rotating Reflecting Surfaces}

\author{Joost de Kleuver \inst{1}, 
        Thomas Bronzwaer \inst{1}, 
        Heino Falcke \inst{1},
        Ramesh Narayan \inst{2,3},
        Yosuke Mizuno \inst{4,5,6}, 
        Oliver Porth \inst{7} \and
        Hector Olivares \inst{8, 1}}

\institute{Department of Astrophysics/IMAPP, Radboud University Nijmegen P.O. Box 9010, 6500 GL Nijmegen, The Netherlands\\
\email{j.dekleuver@astro.ru.nl}
\and
Center for Astrophysics | Harvard \& Smithsonian, 60 Garden Street, Cambridge, MA 02138, USA
\and
Black Hole Initiative at Harvard University, 20 Garden Street, Cambridge, MA 02138, USA 
\and
Tsung-Dao Lee Institute, Shanghai Jiao Tong University, 520 Shengrong Road, Shanghai, 201210, People's Republic of China
\and
School of Physics and Astronomy, Shanghai Jiao Tong University, 800 Dongchuan Road, Shanghai, 200240, People's Republic of China
\and
Institut für Theoretische Physik, Goethe Universit\"at, Max-von-Laue-Str. 1, D-60438 Frankfurt, Germany
\and
Anton Pannekoek Institute for Astronomy, University of Amsterdam, Science Park 904, 1098 XH, Amsterdam, The Netherlands
\and
Departamento de Matem\'{a}tica da Universidade de Aveiro and Centre for Research and Development in Mathematics and Applications (CIDMA), Campus de Santiago, 3810-193 Aveiro, Portugal}
\date{Received ?; accepted ?}

\abstract
{Recently the Event Horizon Telescope observed black holes at event horizon scales for the first time, enabling us to now test the existence of event horizons.} 
{Although event horizons have by definition no observable features, one can look for their non-existence. In that case, it is likely that there is some kind of surface, which like any other surface could absorb (and thermally emit) and/or reflect radiation. In this paper, we study the potential observable features of such rotating reflecting surfaces.}
{We construct a general description of reflecting surfaces in arbitrary spacetimes. This is used to define specific models for static and rotating reflecting surfaces, of which we study the corresponding light paths and synthetic images. This is done by numerical integration of the geodesic equation and by the use of the general relativistic radiative transfer code {\tt RAPTOR}.}
{The reflecting surface creates an infinite set of ring-like features in synthetic images inside the photon ring. There is a central ring in the middle and higher order rings subsequently lie exterior to each other converging to the photon ring. The shape and size of the ring features change only slightly with the radius of the surface $R$, spin $a$ and inclination $i$, resulting in all cases in features inside the `shadow region'.}
{We conclude that rotating reflecting surfaces have clear observable features and that the Event Horizon Telescope is able to observe the difference between reflecting surfaces and an event horizon for high reflectivities. Such reflecting surface models can be excluded, which strengthens the conclusion that the black hole shadow indeed indicates the existence of an event horizon.}


\maketitle

\section{Introduction}
Black holes are a profound prediction of the theory of general relativity (GR) \citep{1915SPAW.......844E}. Although initially seen as purely theoretical objects occurring in highly symmetrical spacetimes \citep{1916AbhKP1916..189S}, the idea was taken more seriously after the discovery that they can be a natural result of gravitational collapse \citep{PhysRev.56.455,PhysRevLett.14.57}. Because of their compactness, deep potential well and possibilities for energy extraction \citep{1971NPhS..229..177P, 1977MNRAS.179..433B}, they became the best candidates for many high energy compact objects in astronomy. It was found that X-ray binaries contain stellar-mass black holes \citep{1972Natur.235..271B, 1972Natur.235...37W, 1986ApJ...308..110M} and that quasars are extremely luminous compact objects best explained by being supermassive black holes (SMBH) \citep{1963Natur.197.1040S, 1969Natur.223..690L}. It now seems that every galaxy contains a dark compact object at its center \citep{1998Natur.395A..14R} and that they may be important for the evolution of their host galaxy \citep{1998AJ....115.2285M, 2012ARA&A..50..455F, 2013ARA&A..51..511K}. Recently, it has even become possible to do precision tests of gravity around black holes with the detection of gravitational waves with LIGO/VIRGO \citep{2016PhRvL.116f1102A,2021PhRvD.103l2002A}, observations of the relativistic orbits of stars around Sagittarius A* (Sgr A*) using GRAVITY \citep{2018A&A...615L..15G, 2020A&A...636L...5G, 2019Sci...365..664D} and horizon scale images of Messier 87* (M87*) and Sgr A* using very long baseline interferometry (VLBI) observations by the Event Horizon Telescope (EHT) \citep{2019ApJ...875L...1E, 2022ApJ...930L..12E}. 

The horizon scale images create an unique way to test gravity \citep{2011JPhCS.283a2030P, 2016ApJ...818..121P, 2017IJMPD..2630001G, 2018GReGr..50...42C, 2019GReGr..51..137P}. The most prominent gravitational effect featured in the images is the black hole shadow \citep{2000ApJ...528L..13F}. It is the region in the image plane bounded by the impact parameters that when traced backward converge to tangent circular orbits \citep{1973ApJ...183..237C, 1979A&A....75..228L}. This is a geometric feature independent of astrophysical effects. In observational images, it shows up as a central brightness depression (CBD). This is produced by the fact that light paths of impact parameters inside the shadow trace backward to the event horizon, therefore having small optical depths compared to impact parameters outside the shadow region \citep{1997A&A...326..419J, 2010ApJ...718..446J, 2019ApJ...885L..33N, 2021arXiv211101123O, 2021ApJ...920..155B, 2022MNRAS.513.1229K}. The radius of the shadow is proportional to the black hole mass and its shape can in some cases give information about the black hole spin, e.g. \citep{PhysRevD.94.084025,2020A&A...636A..94V}.

The accretion around these low luminosity active galactic nuclei is described by the radiatively inefficient advection-dominated flows \citep{1998tbha.conf..148N} and radio jets \citep{1999A&A...342...49F, 2001ApJ...559L..87N}, which are both geometrically thick and optically thin at the observing frequency of 230 GHz. Accurate predictions of the visual appearance of black holes can be made using general relativistic magnetohydrodynamics (GRMHD) models to derive the dynamics of the surrounding plasma and magnetic fields, and general relativistic radiative transfer (GRRT) models to calculate the emission and radiative transfer through curved spacetime. The resulting synthetic images can be compared with observational images of M87* and Sgr A*. This shows that the current observations are in agreement with the sources being Kerr black holes \citep{2019ApJ...875L...5E, 2019ApJ...875L...6E, 2022ApJ...930L..16E, 2022ApJ...930L..17E}.

Despite all this, there remain some fundamental unresolved questions concerning black holes. For one, the Penrose singularity theorem states that in the process of gravitational collapse curvature singularities are formed. The question is whether these are covered by an event horizon (as conjectured by the cosmic censorship conjecture \citep{1969NCimR...1..252P}) or whether they are `naked', i.e. observable. It is thought that quantum theory could prevent these curvature singularities, however, the combination of quantum theory and event horizons leads to the black hole information paradox \citep{1975CMaPh..43..199H} and other related problems \citep{2016RvMP...88a5002H}. This is connected to the prediction in GR called the no-hair theorem that all stationary black hole spacetimes are characterized by only three externally observable classical quantities, namely the mass, charge and angular momentum  \citep{PhysRevLett.11.237, 1967PhRv..164.1776I, 1968CMaPh...8..245I, PhysRev.174.1559, PhysRevLett.26.331, 1972CMaPh..25..152H, 1972PhRvD...5.2419P, 1972PhRvD...5.2439P, 1975PhRvL..34..905R}. In the case of overall charge neutrality, black holes thus should be described by the Kerr metric. A deviation of the Kerr metric together with an observation of an event horizon would then constitute a violation of the no-hair theorem and in extension a violation of GR.

Since the Event Horizon Telescope recently demonstrated the ability to observe black holes at event horizon scales, it has now become possible to look for signs of the existence of event horizons. By definition, event horizons have no direct observational features. To some degree, black hole shadows are their observational feature, however, they could possibly be mimicked by other objects close to an event horizon as well \citep{2020MNRAS.497..521O}. We can look for the observational signs of these latter objects to indicate the possible non-existence of event horizons. If these objects create observational features inside the `shadow region', then that would create a way to distinguish between them and event horizons. If that is the case for some or maybe all such alternatives to event horizons, then that would strengthen the position of black hole shadows as an indicator of the existence of an event horizon.

There are many proposed models for these exotic compact objects \citep{2019LRR....22....4C}. For them to have a shadow, they need to have some sort of light capture cross section. The compactness of these objects can be characterized by the distance of some would-be horizon $\epsilon$. According to Buchdahl's theorem \citep{1959PhRv..116.1027B}, the compactness is bounded under certain conditions. Exotic objects smaller than this bound can be classified by which of these assumptions they violate, such as isotropy for anisotropic stars \citep{2008JDE...245.2243A}. Examples of exotic compact objects possessing some kind of surface are gravastars \citep{Mazur2004}, anisotropic stars, and some boson stars where the bosonic fields interact with photons and other standard model particles. Some examples of alternatives would be naked singularities, wormholes and boson stars with no/weak interaction with photons.

In the case of a surface, the surface could absorb and/or reflect radiation like any other surface. In the case of absorption, the energy should be thermalized and reradiated as thermal emission. This argument is well developed \citep{1998ApJ...492..554N, 2002luml.conf..405N, 2006ApJ...638L..21B, 2007CQGra..24..659B, 2008NewAR..51..733N, 2009ApJ...701.1357B} and has been used to study M87* and Sgr A* using the EHT images \citep{2019ApJ...875L...6E, 2022ApJ...930L..17E}.

Reflecting surfaces have been looked at first in the context of gravitational waves in the form of gravitational wave echoes \citep{2017PhRvD..96h2004A, 2018PhRvD..97l4037W}. In the context of black hole imaging, the argument has only recently been introduced in \citep{2022ApJ...930L..17E} and used to study Sgr A*. They looked at a static spherical surface near where the horizon would be in the Schwarzschild metric. Two assumptions were made: (i) any inward-moving wave vector $k^{\mu}_{\rm i}$ becomes an outward-moving wave vector $k^{\mu}_{\rm f}$ with the radial component $k^r$ reversed and all other components unchanged:
\begin{equation}
    \label{Reflection_law_EHT}
    k^{r}_{\rm f} = - k^{r}_{\rm i} ,
\end{equation}
and (ii) if the intensity of an ingoing ray is $I_{\nu}$, then the intensity of the reflected ray is $A I_{\nu}$, where $A\leq1$ is the albedo of the surface. 
The resulting synthetic images found using such a model contain two new features: a bright central ring and a ring exterior to that both inside the 'shadow region'. For an albedo of $A = 1$, an image with a blur of $15 \mu$as was clearly distinguishable from the image of the EHT and the particular model could be excluded. For lower albedos, the sensitivity is still too low to draw conclusions. This model has been studied further in \citep{2022PhRvD.106h4038C}.

The reflection argument is new and can still be improved upon in many ways. For one, the model does not take rotation into account, despite the fact that the observations indicate the central object having a non-zero spin \citep{2022ApJ...930L..16E}. In this paper, we generalize the model to rotating reflecting surfaces and study the potential observable features of such rotating reflecting surfaces.

The paper is structured as follows. In section \ref{Model}, the mathematical model of reflecting surfaces is described. First, a general model for subsequently static surfaces, moving surfaces in flat spacetime and surfaces in arbitrary spacetimes are discussed and developed. Then, specific models for static spherical surfaces and rotating surfaces are derived. In section \ref{Light_paths} and \ref{Images}, respectively the resulting light paths and synthetic images in these models are explored. In section \ref{Discussion}, the results are discussed and section \ref{Conclusion} is the conclusion.

\section{Models of Reflecting Surfaces}\label{Model}
\subsection{Types of Reflection}
When electromagnetic radiation reaches a boundary between two media, it can be transmitted, absorbed, and/or reflected. The relative amount of incoming radiation that is reflected is given by the albedo $A$, where $A = 0$ means that all is transmitted and/or absorbed and $A = 1$ that all is reflected. Depending on the particular interface, the reflection can be specular, diffuse or a combination of the two. Specular reflection is the mirror-like type of reflection, of which the properties in the case of a plane wave reflecting off a stationary planar surface can easily be derived from the Maxwell equations \citep{jackson_classical_1999}. In that case, the incident and reflected wave vectors form a plane called the plane of incidence, which contains the normal to the surface as well. The angles of the incident and reflected wave vectors with the normal are respectively the angle of incidence $\angle i$ and the angle of reflection $\angle f$. These obey the law of reflection:
\begin{equation}
\label{Reflection_law}
    \angle i = \angle f .
\end{equation}
In addition to that, the frequency and the intensity (up to a factor of the albedo $A$) are conserved.

Diffuse reflection is the type of reflection as seen for example from matte surfaces, where the incoming light is scattered in many different outgoing directions. We will see that given a model of diffuse reflection of a stationary surface, it is straightforward to adjust the reflection law to one describing diffuse reflection. However, the fact that instead of one, there is a continuum of resulting reflected wave vectors, makes it more difficult to do actual calculations with it. Therefore, we will restrict our attention in this paper to specular reflection.

\subsection{Reflection off a Moving Surface in Flat Spacetime}
The reflection of light off an uniformly moving surface was first described by Einstein in his 1905 paper introducing the theory of special relativity \citep{1905AnP...322..891E}. Using a Lorentz transformation he transformed the intensity, angle and frequency of incident light to the frame of the surface. There he used the known reflection behavior for a stationary surface as described above. Finally, he transformed the reflected quantities back to the observer frame using the inverse Lorentz transformation. The surprising result obtained was that the angles of incidence and reflection no longer have to be equal, contrary to Eq. \ref{Reflection_law}. The same is true for the incident and reflected intensity and frequency. All three quantities now depend on the velocity and orientation of the surface. This result has since then been well established, experimentally tested and rederived many times \citep{1989SvPhU..32..813B, 2004AmJPh..72.1316G, 2012AmJPh..80..680G}. 

One computationally useful way to describe the reflection is in terms of the incident and reflected wave vectors $k_{\rm i}$ and $k_{\rm f}$, which are essentially the four-momenta of incident and reflected photons. The wave vectors are null vectors, although all following results will be true as well for the four-momenta of massive particles scattering in an elastic collision from the surface without friction. The reflection law is given by:
\begin{equation}
    k_{\rm f} = \Lambda^{-1} \, R \, \Lambda \, k_{\rm i} ,
\end{equation}
where $\Lambda$ is the Lorentz transformation from the observer frame to the stationary frame of the surface and $R$ is the reflection transformation in the surface frame. Since the Lorentz group contains rotations, one can choose without loss of generality the Lorentz transformation such that the reflection transformation has the form $R = {\rm diag}(1, -1, 1, 1)$.

An alternative and maybe more insightful way for building intuition, is to look at reflection using the Huygens-Fresnel principle. This is the formalism stating that every point on a wavefront is a source of spherical waves, which mutually interfere to create a new wavefront. When a wavefront approaches a stationary planar surface at an angle, points on the wavefront reach the surface at different times, resulting in spherical secondary waveforms of different radii, combining into a new wavefront that moves outward at an equal reflected angle. Now when the surface moves in the direction of its normal, the scattering centers will lie on a different line than before, inclined with respect to the surface, which results in a different reflection angle. The difference in frequency and intensity can be explained by the fact that the waves are in some sense pressed together or stretched by the movement of the surface.

In this formalism, it can be seen that the only aspect of the surface that matters for the reflection is its position as a whole. The motion of the surface tangential to itself, so motion that keeps the position of the surface as a whole invariant, does not have an influence on the reflection behavior. This can be seen using the wave vector description as well. When a surface has a normal in the x-direction and moves in the y- or z-direction, the matrices $\Lambda$ and $R$ commute. This results in the Lorentz transformations canceling each other leaving only the reflection transformation of the stationary frame. From this can be concluded that all relevant information for reflection is contained in the hypersurface spanned by the worldlines of the surface.

The result above can be generalized to curved and/or non-uniformly moving surfaces. Using the Huygens-Fresnel principle, it is clear that the resulting wavefront will generally not be a plane, if it is even still possible to talk about a clear wavefront. However, the reflection can locally be approximated by the reflection of a planar surface if the surface is spatially flat with respect to the wavelength ($\lambda \ll R$, where $R$ is the local radius of curvature) and `temporally flat' with respect to the frequency ($\nu \gg \frac{a}{2c}$, where $a$ is the acceleration).

\subsection{Reflection in Arbitrary Spacetimes}
Now we focus our attention towards reflection of a surface in a general Lorentzian spacetime $(M,g_{\mu \nu})$. A physical surface can be described by a time-like hypersurface $S \subset M$. Based on the discussion above, this is all the information we need to describe reflection. To be able to locally approximate the hypersurface by a plane, the surface should again be `flat enough' in a normal coordinates frame in a neighborhood of the point of reflection.

A way to define the time-like hypersurface $S$ is by defining a smooth time-like vector field $U \in TS \subset TM$ describing the velocity of each point of the surface. The integral curves of $U$ then describe the worldlines of spatial points on the surface.

The paths of light through spacetime are described by the geodesic equation:
\begin{eqnarray}
    \label{Geodesic_equation_1}
    \frac{{\rm d}x^{\rho}}{{\rm d} \lambda} &=& k^{\rho} ,\\
     \label{Geodesic_equation_2}
    \frac{{\rm d}k^{\rho}}{{\rm d} \lambda} &=& - \Gamma^{\rho}_{\mu \nu} k^{\mu} k^{\nu},
\end{eqnarray}
where $x^{\rho}(\lambda)$ is the spacetime coordinate describing the light path, $k^{\rho} (\lambda)$ the corresponding wave vector, $\lambda$ an affine parameter parametrizing the path and $\Gamma^{\rho}_{\mu \nu}$ the Christoffel symbol which is a function of the metric $g_{\mu \nu}$:
\begin{equation}
    \Gamma^{\rho}_{\mu \nu} = \frac{1}{2} g^{\rho \alpha} \left[ \partial_{\mu} g_{\nu \alpha} + \partial_{\nu} g_{\alpha \mu} - \partial_{\alpha} g_{\mu \nu} \right] .
\end{equation}
An incident light ray can be chosen by choosing initial conditions $x^{\rho}(0)$ and $k^{\rho}(0)$. By integrating the geodesic equation, its path can be calculated until it intersects the hypersurface $S$. We will call the intersection point $p$.

The reflection in this point can be calculated by transforming to a local reference frame, since the reflection in flat spacetime is known. This can be done by transforming to a normal coordinates frame of that point.

Another equivalent way is by defining a tetrad $e^{\mu}_{(i)}$. This is an orthonormal basis of the tangent space $T_pM$ containing one time-like and three space-like vectors, which for any choice corresponds to the tangent space of a local frame at that point. For simplicity in the calculations, we choose the tetrad in the following way. We choose $e_{(0)} \in TS_p$ to be a time-like unit tangent vector, for example the velocity vector $U(p)$. For $e_{(2)}$, $e_{(3)} \in TS_p$, we choose orthonormal space-like  tangent vectors orthonormal to $e_{(0)}$ and each other. Finally, we choose for $e_{(1)} = N(p) \in TS^{\perp}_{p}$ a unit normal vector to the surface. This unit normal vector is unique up to a sign and can be chosen uniquely in the direction of the incident and reflected waves. The transformation matrix from the tetrad basis to the coordinate basis is then given by:
\begin{equation}
    e^{\mu}_{(i)} = 
    \begin{pmatrix}
    U^{\mu} & N^{\mu} & e^{\mu}_{(2)} & e^{\mu}_{(3)}
    \end{pmatrix} .
\end{equation}
The outgoing wave vector $k_{\rm f}$ can now be derived using
\begin{equation}
    \label{Reflection_law_general}
    k_{\rm f}^{\nu} = e^{\nu}_{(i)} \: R^{(i)}_{(j)} \: e^{(j)}_{\mu} \: k_{\rm i}^{\mu} ,
\end{equation}
where $e^{\nu}_{(i)}$ and $e^{(j)}_{\mu}$ are respectively the transformation matrices from the tetrad basis to the coordinate basis and its inverse, and $R^{(i)}_{(j)} = {\rm diag}(1, -1, 1, 1)$ is the reflection transformation in the tetrad basis.

The path of the reflected light can now be calculated by integration of the geodesic equation starting from $p$ and using the reflected wave vector $k_{\rm f}$.

The model can be adjusted to describe diffuse reflection by replacing the reflection transformation $ R^{(i)}_{(j)}$ with a distribution of transformations describing diffuse reflection of a stationary surface. This will result in a distribution of reflected wave vectors of which all geodesics have to be calculated. Here possibly new interesting behaviour can occur such as light rays that bounce multiple times on the surface before they escape.

\subsection{Reflection off a Static Spherical Surface}
We are now able to derive a model for the reflection of light off a static spherical surface. Assuming a spherically symmetric mass distribution inside the surface, according to Birkhoff's theorem \citep{Birkhoff} the metric outside the surface is given by the Schwarzschild metric. In geometrized units, meaning $c = G = 1$, where $c$ is the speed of light and $G$ is the gravitational constant, the metric is given by:
\begin{equation}
\label{Schwarzschild_metric}
    {\rm d}s^2 = - \left(1 - \frac{2 M}{r} \right) {\rm d}t^2 + \left(1 - \frac{2 M}{r} \right)^{-1}  {\rm d}r^2 + r^2 \left(  {\rm d}\theta ^2 + \sin^2 \theta  {\rm d}\phi ^2 \right) ,
\end{equation}
where ${\rm d}s^2$ is the line element and $M$ is the mass. The hypersurface describing the spherical surface of radius $R > 2M$ through time is given by $S = \{x\,^{\rho} \in M \,|\, x^r = R \, \}$. The velocity field is given by $U =  \left(1 - \frac{2M}{R} \right)^{-1/2} \frac{\partial}{\partial t}$.

Now it can easily be seen that a possible tetrad basis is given by:
\begin{equation}
    e^{\mu}_{(i)} = 
    \begin{pmatrix}
    \left(1 - \frac{2M}{R} \right)^{-1/2} & 0 & 0 & 0\\
    0 & \left(1 - \frac{2M}{R} \right)^{1/2} & 0 & 0\\
    0 & 0 & R^{-1} & 0\\
    0 & 0 & 0 & \left(R\ |\sin \theta| \right)^{-1}
    \end{pmatrix} .
\end{equation}
This matrix commutes with the reflection transformation matrix $R^{(i)}_{(j)}$. It, therefore, cancels against its inverse, leaving only the reflection transformation. The resulting reflection law is:
\begin{equation}
\label{Reflection_law_static}
    k^{r}_{\rm f} = - k^{r}_{\rm i} .
\end{equation}

\subsection{Reflection off a Rotating Surface}
Finally, we discuss the reflection of light off a rotating surface. The difficulty here is that there is no clear choice for a metric and hypersurface. One has to solve for the corresponding spacetime and in general this would not be given by the Kerr metric. Since the already available GRMHD simulations use the Kerr metric, it is practical to use the Kerr metric. This is justified by the fact that this is a first qualitative exploration of the appearance of a rotating reflecting surface and that the Kerr metric to some degree approximates the metric of a rotating surface. In future research, it would be interesting to look at metrics for specific mass distributions. The Kerr metric in geometrized units is given by:
\begin{multline}
    {\rm d}s^2 = - \left ( 1 - \frac{2 M r}{\rho^2} \right) {\rm d}t^2 - \frac{4 M a r \sin^2 \theta}{\rho^2} {\rm d}t {\rm d}\phi + \frac{\rho^2}{\Delta} {\rm d}r^2 \\
    + \rho^2 {\rm d}\theta^2 + \frac{\sin^2 \theta}{\rho^2} \left[ (r^2 + a^2)^2 - a^2 \Delta \sin^2 \theta \right] {\rm d}\phi^2 ,
    \label{Kerr_metric}
\end{multline}
where
\begin{eqnarray}
 \label{Kerr variabelen}
    \Delta &=& r^2 - 2 M r + a^2 , \\
    \rho^2 &=& r^2 + a^2 \cos^2 \theta ,
\end{eqnarray}
and $a$ is the spin given by
\begin{equation}
    a = \frac{J}{M} ,
\end{equation}
where $J$ is the angular momentum of the spacetime. By expressing the spin $a$ in units of the black hole mass $M$, one finds the dimensionless spin parameter, which will be used to describe the value of the spin throughout this paper.

As surface we use the constant radius surface with radius $R > r_+$, where $r_+ = 1 + \sqrt{1 - a^2}$ is the event horizon radius in units of black hole mass $M$. The surface is described by the hypersurface $S = \{x\,^{\rho} \in M \,|\, x^r = R \, \}$. We make this choice because it is the natural generalization of a spherical surface, it converges to the event horizon when taking the limit of the radius $R \downarrow r_+$ and it approximates a rigidly rotating zero-angular momentum observer (ZAMO) surface \citep{2014PhRvD..90l4010F} when choosing as velocity that of a ZAMO observer.

For constructing a tetrad, it is most convenient to define a velocity field $U$. As discussed earlier, it does not matter which time-like velocity field tangent to the surface we choose. This means that the resulting reflection law does not depend on the rotation of the surface, which makes the result more general. The only effect of rotation on the reflected light is through the frame-dragging of the spacetime.

We choose the velocity of a ZAMO observer: $U = \beta \left( \frac{\partial}{\partial t} + \omega \frac{\partial}{\partial \phi} \right)$, where $\omega = \frac{{\rm d} \phi}{{\rm d}t} = - g_{t \phi}/g_{\phi \phi}$ is the angular velocity and $\beta = \sqrt{- g^{t t}}$ is a normalization factor.

A possible tetrad is then given by:
\begin{equation}
    e^{\mu}_{(i)} = 
    \begin{pmatrix}
    \sqrt{-g^{t t}} & 0 & 0 & 0\\
    0 & g_{r r}^{-1/2} & 0 & 0\\
    0 & 0 & g_{\theta \theta}^{-1/2} & 0\\
    \sqrt{- g^{t t}} \ \omega & 0 & 0 & g_{\phi \phi}^{-1/2}
    \end{pmatrix} .
\end{equation}
This matrix commutes with the reflection transformation matrix $R^{(i)}_{(j)}$, just like in the static case. It, therefore, cancels against its inverse and results in the reflection law:
\begin{equation}
\label{Kerr_reflection}
    k^{r}_{\rm f} = - k^{r}_{\rm i} .
\end{equation}
It is important to note that this equation is dependent on the coordinates and generally is much more complex in other coordinates than Boyer-Lindquist coordinates. When using other coordinates, the reflection law has to be carefully transformed to those coordinates. Another possibility is to transform the incident wave vector to Boyer-Lindquist coordinates, use the simple reflection law there and then transform the resulting reflected wave vector back.

A result of the reflection law is that there will be no change in the frequency and intensity in the reflection. The frequency does, however, change along the path toward and away from the surface because of the gravitational red and blue shift. In the Schwarzschild and Kerr metric, these effects perfectly cancel, meaning that ingoing and reflected outgoing light have the same frequency independent of the radius of the surface $R$. This also means that a smooth radial transition from a surface $R > r_+$ to an event horizon would result in an abrupt change from seeing reflected light at a constant frequency to seeing no reflected light at all. This on-off type effect allows us to distinguish between reflecting surfaces and event horizons independent of the radius $R$, even for radii infinitesimally close to $r_+$.

A remarkable result now is that the photons do not gain energy from the rotating surface. This remains true for other surfaces that are stationary as well. Take for example a rigidly rotating ZAMO surface. It is almost a constant radius surface, but does bulge out a bit around the equator. This will result in slightly different reflection angles. Despite this, the frequency will still be unchanged because of the stationarity of the hypersurface.

There would be a change in frequency if the surface would move radially. This is for example what would happen for light reflected off a collapsing ball of dust. The frequency of the reflected light would decrease indefinitely with the increasing in fall velocity.

Finally, we can compare the wave-like reflection with light scattering off some sort of atmosphere. One could for example consider inverse Compton scattering on atmospheric particles. The derivation of the scattering of a single particle is somewhat similar to the derivation of the reflection law above, in that a Lorentz transformation is done to the rest frame of the particle, then the scattering is calculated, and finally an inverse Lorentz transformation is done. Yet, in this case one finds that the velocity of the atmospheric particle does matter, despite the fact that the velocity of the particles is in the tangential direction of the surface. The energy and frequency of the light will increase by about a factor $\gamma^2 = \left(1 - v^2/c^2\right)^{-1}$, which will change the angle of reflection from the description above. However, if we choose as a surface a ZAMO surface, the atmosphere would have zero velocity in its rest frame, and the frequency would still be conserved. So although the 'reflecting' atmosphere has a relativistic velocity with respect to a distant observer, the velocity is gone when the light ray reaches the ZAMO frame of the surface. In all cases, it can be expected that in a realistic atmospheric model, the resulting reflected light will become somewhat diffuse and span a broader range of frequencies.

\begin{figure*}
   \centering
   \includegraphics[width=\textwidth]{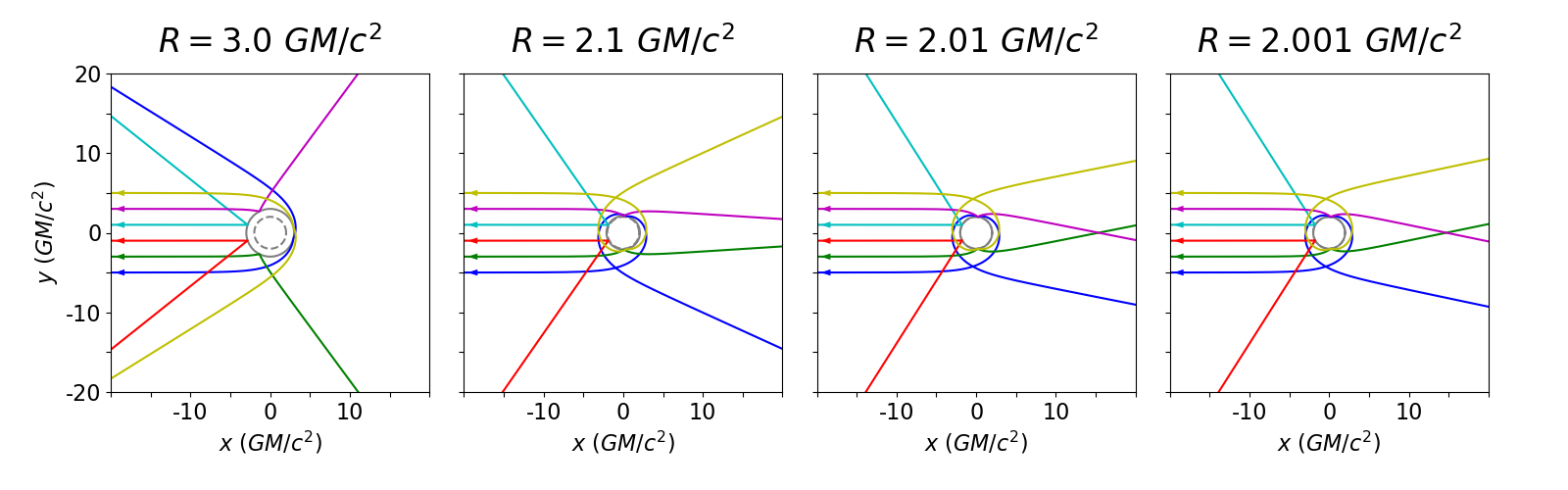}
   \caption{Null geodesics in the equatorial plane of static spherical reflecting surfaces of different radii $R$. The impact parameter $b$ ranges from $-5$ to $5$ $GM/c^2$. The solid grey line indicates the surface and the dashed grey line the location where the event horizon would have been.}
    \label{Geodesics_static}%
    \end{figure*}
 \begin{figure}
   \centering
   \includegraphics[width=9cm]{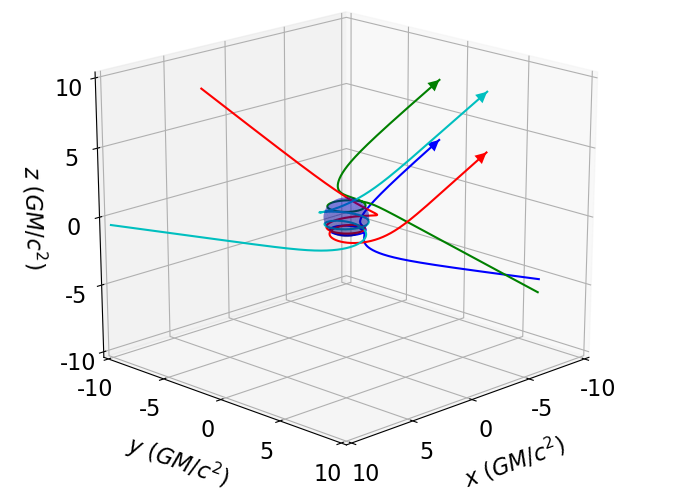}
   \caption{Null geodesics around a rotating reflecting surface with spin $a = 0.94$ and surface radius $R = 1.01 r_+$, where $r_+$ is the radius an event horizon with the corresponding spin would have had. The geodesics are directed towards a far-away observer in the negative x-direction having an inclination $i = 60 \, ^{\circ}$. The image plane parameters $(\alpha, \beta)$ are $(2,2)$, $(2,-2)$, $(-2,2)$ and $(-2,-2)$ $GM/c^2$.}
    \label{Geodesics_3d}%
    \end{figure}
    
\section{Light Paths}\label{Light_paths}
\subsection{Method}\label{Geodesics_method}
To study light paths in the models discussed above, we modified a stripped-down version of the general relativistic ray-tracing code {\tt RAPTOR} \citep{Bronzwaer2018RAPTORIT} that calculates null geodesics. It does this by numerical integration of the geodesic equation, Eq. \ref{Geodesic_equation_1} and Eq. \ref{Geodesic_equation_2}, using the fourth order Runge-Kutta method (RK4). As metric we use the Kerr metric in Boyer-Lindquist coordinates, Eq. \ref{Kerr_metric}, with spin $a$ as a free parameter. For spin $a = 0$, this corresponds to the Schwarzschild metric in Schwarzschild coordinates, Eq. \ref{Schwarzschild_metric}.

The initial position $x_0$ and wave vector $k_0$ are chosen far-away from the spatial origin in the direction of a line-of-sight. The line starts at the spatial origin and can, for example, be chosen along the $x$-axis or along a line with an inclination angle $i$ with respect to the spin axis. The initial position $x_0$ is then chosen far-away along this line with an offset perpendicular to the line specified by the impact parameter(s). In two dimensions, we will describe the offset with the impact parameter $b$. In three dimensions, this is described by image plane parameters $\alpha$ and $\beta$, where $\alpha$ is in the direction perpendicular to the line-of-sight aligned with the spin axis and $\beta$ in the perpendicular direction. The spatial part of the wave vector $k_0$ is then chosen parallel to the line-of-sight and away from the spatial origin. The time component of the wave vector is then determined by demanding the wave vector to be null.

Integration starting from the initial conditions can be done forward or backward, so respectively in the direction of the wave vector or in the opposite direction. The latter option corresponds to calculating the path a light-ray followed to reach the position in the camera. This option is what is used in {\tt RAPTOR} and is what we will use here as well.

To determine the paths of light rays around a reflecting surface with radius $R$, we have to take the possibility of reflection into account. This is done by checking in each integration step if the radial coordinate has decreased below the radius $R$, since that indicates that the reflecting surface was intersected. If that is the case, the reflection procedure is started.

First, the intersection point and its corresponding wave vector are determined. The path crossed the surface in the last integration step, so there should be a value of the integration step size $\lambda$ between $0$ and the used integration step size $\lambda_{\rm 0}$ such that the resulting radial coordinate equals the radius of the surface $R$. The bisection method is used to solve $r (\lambda) - R = 0$ for step size $\lambda$, where $r (\lambda)$ is the radial coordinate found in the last integration step when using step size $\lambda$. In each iteration step, the integration step is calculated over again using the new step size. The resulting value for the step size $\lambda$ can be used to calculate the integration step one final time to determine the position and the corresponding wave vector at the intersection point.

After this, the reflection is calculated. Since we are using Boyer-Lindquist coordinates, this is simply done by flipping the radial component of the wave vector, as described in Eq. \ref{Kerr_reflection}. This works for both forward and backward integration.

Finally, the geodesic can be integrated further from the reflection point using the new wave vector. In principle it can reflect again from the surface if the radial coordinate becomes smaller than the surface radius $R$ again. This could give some problems since the calculated radial coordinate of the intersection point can be smaller than the surface radius $R$. To prevent unphysical behavior, the reflection procedure can only start again after the geodesic has been above the surface radius $R$ again.

\subsection{Static Spherical Surfaces}
We will first look at null geodesics around a static spherical surface, so at spin $a = 0$. In that case, the geodesics stay in an orbital plane, which we can choose without loss of generality to be the equatorial plane. We choose the direction of the observer to be in the negative $x$-direction and use impact parameter $b$ to specify geodesics. Null geodesics around static spherical reflecting surfaces of different radii $R$ are shown in Fig. \ref{Geodesics_static}.

The first thing to notice, is that there is symmetry in the $x$-axis. This is because of the absence of the frame-dragging effect and is a logical result of the spherical symmetry. Geodesics with impact parameters $\pm b$ will therefore describe the same behavior.

On the left plot, geodesics are shown around a surface of radius $R = 3.0$ $GM/c^2$, which corresponds to the photon sphere. This means that geodesics with impact parameters smaller than the photon capture radius $b = 3\sqrt{3}$ $GM/c^2$ reach the surface and are reflected. Geodesics with impact parameter larger than the photon capture radius do not reach the surface nor the photon sphere and are thus the same as geodesics with the same impact parameter around a static black hole. This behavior is true for all surface radii smaller than the radius of the photon sphere.

The geodesics for $b = \pm 1$ on the left plot look like the reflection off a reflecting ball in flat spacetime; straight lines in and out. This is expected for geodesics with small impact parameters $b$ reflecting of surfaces with large radii $R$, since in these cases the geodesics move mostly radially and do not or only briefly go through a region of strong gravity. Indeed, for larger impact parameters $b = \pm 3$ and $b = \pm 5$, the geodesics are increasingly more curved. This results for $b = \pm 3$ in a smaller and for $b = \pm 5$ in a larger deflection angle than that of a ball in flat spacetime. Increasing the impact parameter further would result in increasingly larger deflection angles, where the geodesic will orbit the surface many times. The deflection angle will diverge when going to the photon capture radius $b = 3\sqrt{3}$ $GM/c^2$, which corresponds to a geodesic that asymptotically converges to a circular photon orbit.

For smaller radii, the deflection angles increasingly become larger as can be seen in the other plots in Fig. \ref{Geodesics_static}. This is a result of the fact that the geodesics have to travel farther to reach the surface. It can be seen that even the geodesic with $b = \pm 1$ deflects a bit now as expected.

The deflection angle seems to converge when decreasing the radius, with smaller impact parameters nearing their final deflection angle faster than higher impact parameters. The convergence of the deflection angle is expected, since there is an upper limit on the deflection angle set by two times the angular distance to reach the event horizon radius $2GM/c^2$. The lower impact parameters converge faster because their corresponding geodesics have to travel a smaller angular distance.

\begin{figure*}
   \centering
   \includegraphics[width=\textwidth]{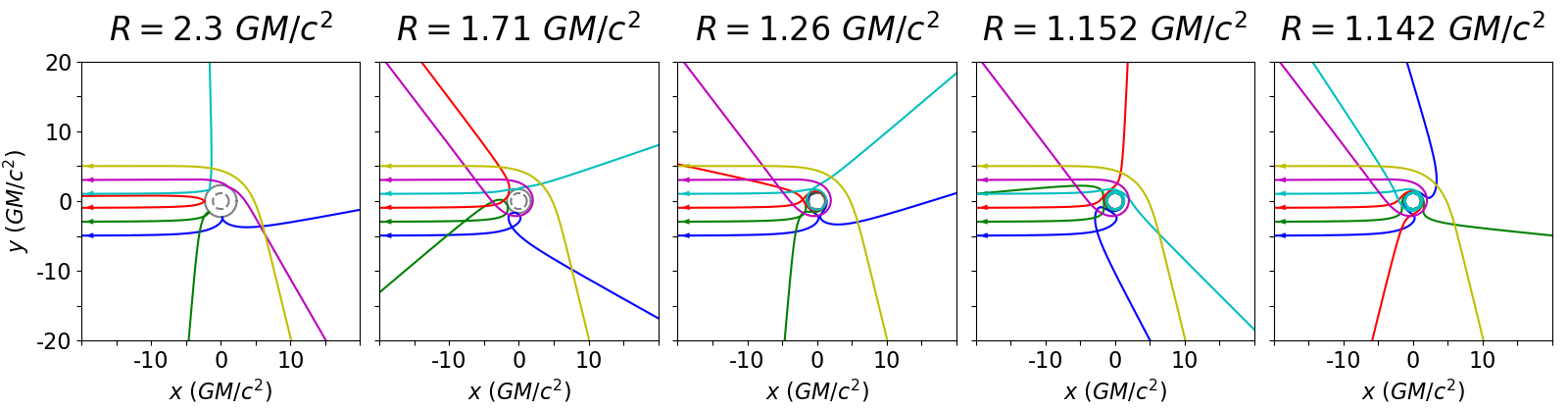}
   \caption{Null geodesics in the equatorial plane of rotating reflecting surfaces of different radii $R$ and with spin $a = 0.99$. The radii correspond from left to right to $2.0$, $1.5$, $1.1$, $1.01$ and $1.001$ times the radius $r_+$ an event horizon would have had. The impact parameter $b$ ranges from $-5$ to $5$ $GM/c^2$. The solid grey line indicates the surface and the dashed grey line the location where the event horizon would have been.}
    \label{Geodesics_rotating_radius}%
    \end{figure*}
\begin{figure*}
   \centering
   \includegraphics[width=\textwidth]{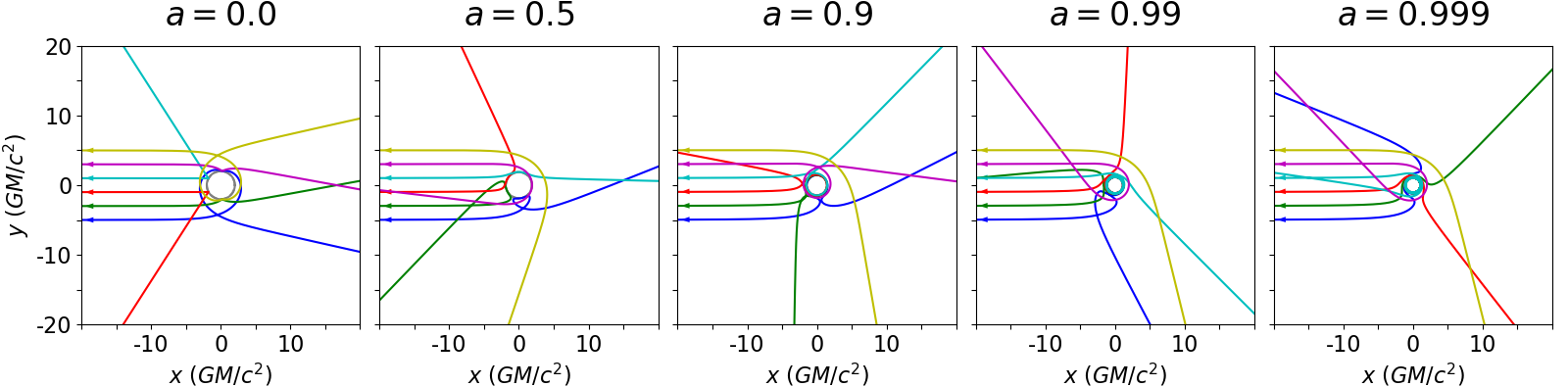}
   \caption{Null geodesics in the equatorial plane of rotating reflecting surfaces of different spin $a$ and with surface radius $R = 1.01 r_+$, where $r_+$ is the radius an event horizon with the corresponding spin would have had. The impact parameter $b$ ranges from $-5$ to $5$ $GM/c^2$. The solid grey line indicates the surface and effectively coincides with the location where the event horizon would have been because of the high spin.}
    \label{Geodesics_rotating_spin}%
    \end{figure*}

\subsection{Rotating Surfaces}
Null geodesics around a rotating reflecting surface are shown in Fig. \ref{Geodesics_3d}. They are directed towards a far-away observer with inclination $i = 60 ^{\circ}$. The frame-dragging effect causes the geodesics to get dragged along the spin axis resulting in spiraling behavior.
Unlike the static case, the geodesics generally do not stay in an orbital plane. The exception are geodesics in the equatorial plane. To get a better understanding of the behavior of the geodesics, we take a look at these. Geodesics in the equatorial plane around a reflecting surface with different radii $R$ are shown in Fig. \ref{Geodesics_rotating_radius} and with different spin $a$ in Fig. \ref{Geodesics_rotating_spin}.

First of all, it can directly be seen that in the rotating case, there is no symmetry in the $x$-axis anymore. This is because moving clockwise and anticlockwise is different due to the frame-dragging effect. Geodesics with impact parameters $\pm b$ will therefore not describe the same behavior anymore.

In the second plot in Fig. \ref{Geodesics_rotating_radius}, one can see that the line with impact parameter $b = - 5$ has some interesting behavior: it makes a loop. It starts on the right moving anti-clockwise, then switches direction to moving clockwise, reflects and finally switches direction again to finally move anti-clockwise again. At first sight, this seems like strange behavior, since it changes to the anti-clockwise direction after being reflected in the clockwise direction. However, this is a misconception caused by the coordinates. In the ZAMO frame of the reflection point, the reflection happens in the anti-clockwise direction. In the coordinate system of the static observer, it only looks like the reflection is in the other direction, but one has to take the frame-dragging into account to fully appreciate what happens. When the geodesic moves inward, it is dragged along with the spacetime more and more. At a certain moment, it is dragged along more in the anti-clockwise direction than it can move clockwise, which results in a total anti-clockwise motion in the coordinate frame. This is also the case directly after the reflection. Then when it moves further away from the surface, the frame-dragging effect decreases again until the geodesics move clockwise in the coordinate frame again.

The photon orbit changes with spin. In the rotating case, there are two, one prograde ($+$) and one retrograde ($-$). Their radius $r_{\rm ph}^{\pm}$ is given by \citep{1972ApJ...178..347B}:
\begin{equation}
    r_{\rm ph}^{\pm} = 2 \frac{GM}{c^2} \left[ 1 + \cos \left( \frac{2}{3} \arccos \left( \mp|a| \right) \right) \right] .
\end{equation}
The corresponding shadow radii in the equatorial plane are a bit larger and will for the following discussion be denoted by $b^{\pm}$. For $R < r_{\rm ph}^{+} \leq r_{\rm ph}^{-}$, the geodesics corresponding to impact parameters $b \in \left(-b^-, b^+ \right)$ will reach the surface and be reflected. Geodesics corresponding to impact parameters $b$ outside $[-b^-,b^+]$ do not reach the surface nor the photon orbit and are thus the same as geodesics with the same parameter around a rotating black hole with spin $a$.

For $r_{\rm ph}^{+} < R < r_{\rm ph}^{-}$, the surface extends for prograde orbits above the photon radius. This means that prograde orbits are different from those around a black hole and that geodesics corresponding to impact parameters $b \in \left(-b^-, b_R^+ \right)$ reach the surface, where $b_R^+ > b^+$ is the new boundary impact parameter. The transition between these two cases can be seen in Fig. \ref{Geodesics_rotating_radius} for impact parameter $b = 3$. In the first plot, it reflects from the surface, but in the later plots, it does not.

For differing spin, it is important to note that the radius of the horizon $r_+$ and therefore of the surface radius $R = 1.01 r_+$ as well decrease with spin. The photon radius for prograde orbits $r_{\rm ph}^{+}$ will also decrease, which can be seen for impact parameters $b = 5$ and $b = 3$. In the first plot, they reflect from the surface, but in the later plots, they do not. For retrograde orbits, the photon radius increases, resulting in a larger range of impact parameters falling to the surface. Finally, when increasing the spin to $1$ both the radius of the event horizon $r_+$ and the radius of the prograde photon orbit $r^{\rm ph}_+$ will converge to the same value. This means that for spins very close to $1$, the radius of the surface $R$ will become higher than the prograde photon orbit $r_{\rm ph}^{+}$ ($R = 1.01 r_+ \rightarrow 1.01 r_{\rm ph}^{+}$, giving $R > r_{\rm ph}^{+}$). So for these spin values there will be prograde geodesics, that are reflected outside the photon sphere.

Just as in the static case, one would expect the deflection angles to converge when decreasing the radius. This only seems to happen for the geodesics with impact parameters $b = 5$ and $b = 3$ because these do not reach the surface for smaller radii. It is expected that it takes very small radii to converge, because the frame-dragging causes the geodesics to travel long angular distances as can be seen in the plots. There should, however, be geodesics with small deflection angles that consequently should converge already at relatively large radii. Since the angular deflection continuously goes from $-\infty$ for $b^-$ towards $+\infty$ for $b^+$, there should be a $b \in \left( b^-, b^+ \right)$ such that the deflection angle is $0$.

For the spin, in Fig. \ref{Geodesics_rotating_spin}, there is no clear convergent behavior with increasing spin visible. Just like in the case with decreasing radius, one would expect convergence of the geodesics when increasing the spin to $a = 1$, since it changes continuously to $1$ and there is a specific solution for $a = 1$. Again, the convergence will be faster for impact parameters with small deflection angles.

Coming back to the general case for orbits outside the equatorial plane, similar behaviour is expected, but now with orbital motion in the $\theta$-direction as well. Just as before, geodesics will fall to the surface when going below the photon orbit and be reflected. For large surface radii, the surface can (partly) stick above the photon orbits, resulting in reflected geodesics there as well. We also expect similar converging behaviour for surface radius $R$ to event horizon radius $r_+$ and spin $a$ going to $1$.

\section{Synthetic Images}\label{Images}
\subsection{Method}
To make synthetic images, we use a modified version of the general radiative transfer code {\tt RAPTOR} \citep{Bronzwaer2018RAPTORIT}. It calculates null geodesics and integrates the relativistic radiative transfer equation along them. To use it for synthetic images of reflecting surfaces, we modified the calculation of the geodesics in the same way as discussed in section \ref{Geodesics_method}. The radiative transfer is subsequently integrated along the reflected geodesic.

The only difference in the calculation of the geodesics compared to section \ref{Geodesics_method}, is that we use modified Kerr-Schild coordinates as defined for the GRMHD code {\tt BHAC} \citep{Porth_2017} instead of Boyer-Lindquist coordinates. This means that the reflection law generally does not work anymore, since Eq. \ref{Kerr_reflection} is in Boyer-Lindquist coordinates. To solve this, when calculating the reflection, the wave vector is transformed to Boyer-Lindquist coordinates, reflected using Eq. \ref{Kerr_reflection} and then transformed back.

The synthetic images have a field of view of $30 \times 30$ $GM/c^2$ with resolution of $1000 \times 1000$ pixels for the analytic Keplerian torus model and $256 \times 256$ pixels for the GRMHD models. An observing frequency of $230$ GHz is used, at which the gas is optically thin. 

\subsection{Analytic Keplerian Torus Model}
\begin{figure*}
   \centering
   \includegraphics[width=\textwidth]{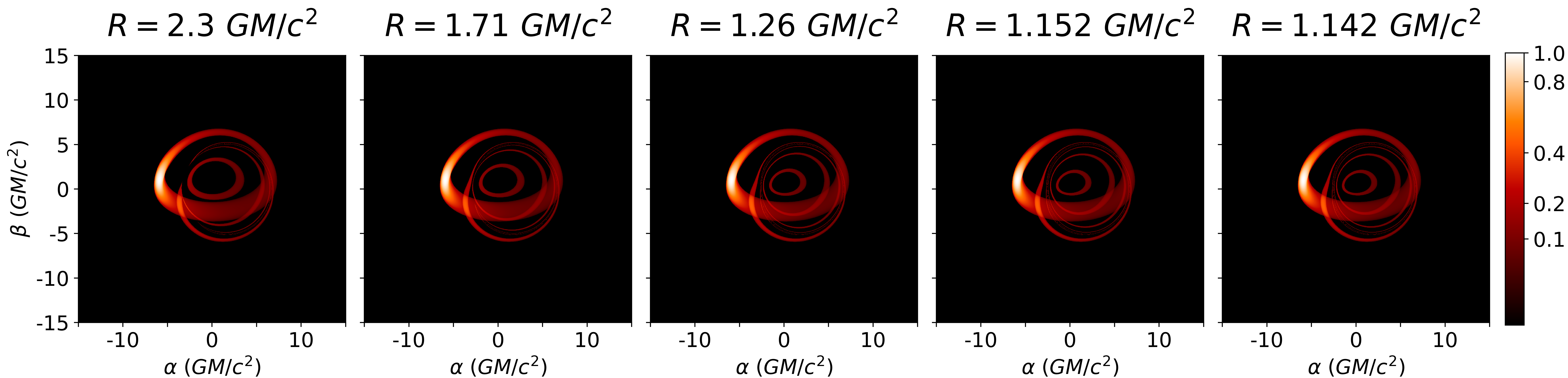}
   \caption{Synthetic images of rotating reflecting surfaces with different radii $R$, with spin $a = 0.99$, inclination $i = 60 ^{\circ}$ and surrounded by an analytic Keplerian torus. The radii correspond from left to right to $2.0$, $1.5$, $1.1$, $1.01$ and $1.001$ times the radius $r_+$ an event horizon would have had.}
    \label{Rad_radius}%
    \end{figure*}
\begin{figure*}
   \centering
   \includegraphics[width=\textwidth]{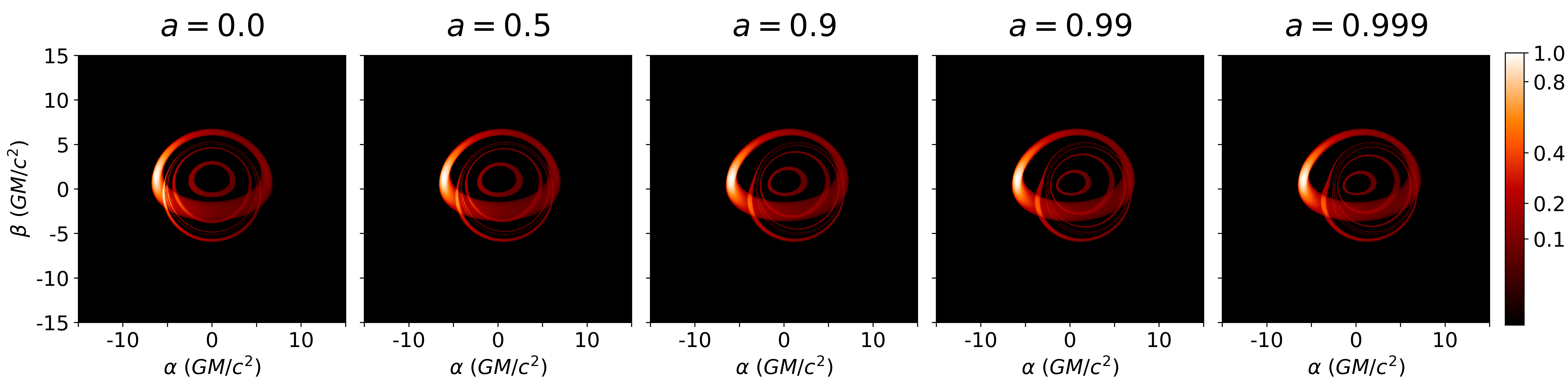}
   \caption{Synthetic images of rotating reflecting surfaces with different spin $a$, inclination $i = 60 ^{\circ}$ and surrounded by an analytic Keplerian torus. The surface radius is $R = 1.01 r_+$, where $r_+$ is the radius an event horizon with the corresponding spin would have had.}
    \label{Rad_spin}%
    \end{figure*}
\begin{figure*}
   \centering
   \includegraphics[width=\textwidth]{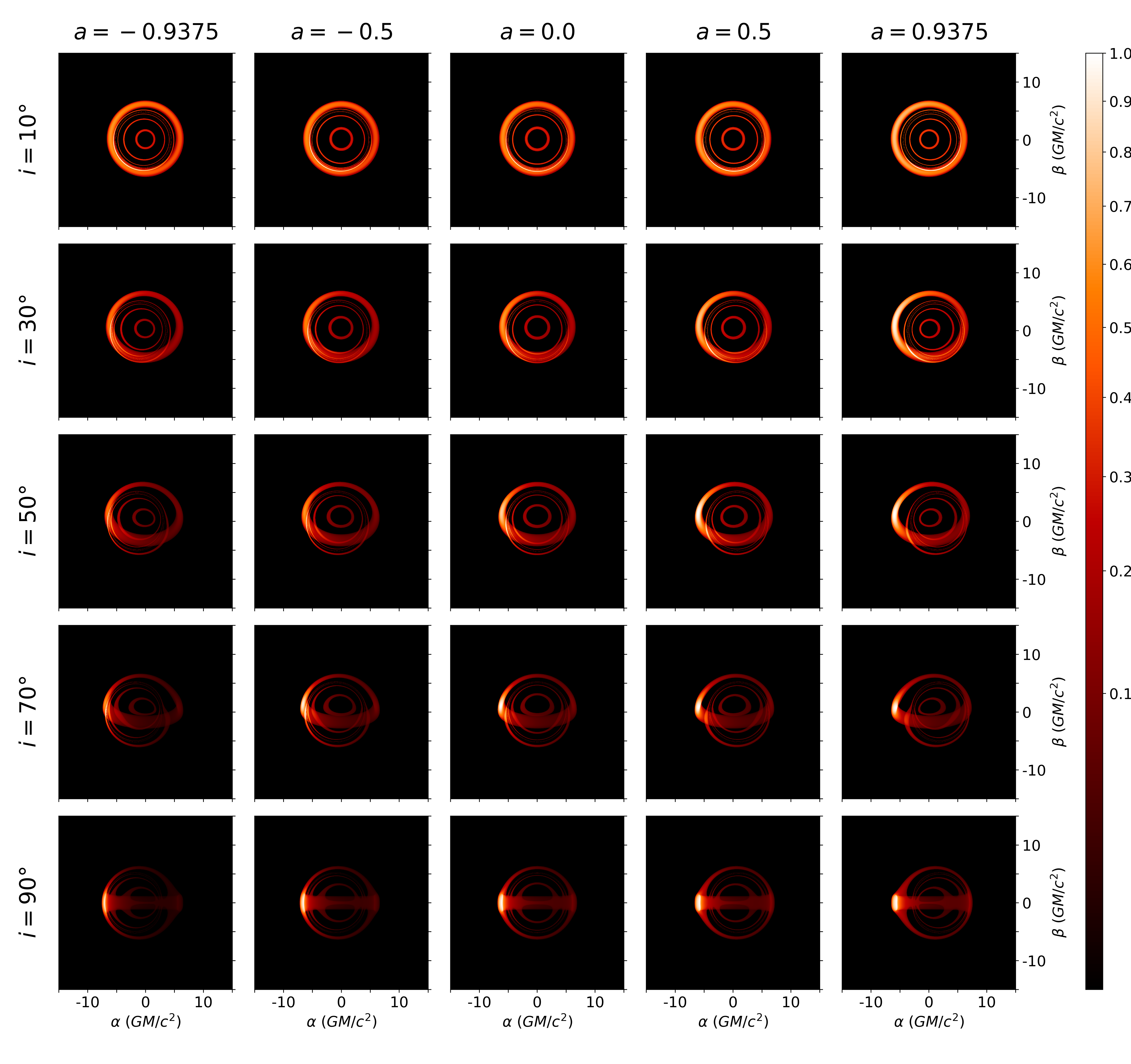}
   \caption{Synthetic images of rotating reflecting surfaces with different spin $a$ and inclination $i$, surrounded by an analytic Keplerian torus.  The surface radius is $R = 1.01 r_+$, where $r_+$ is the radius an event horizon with the corresponding spin would have had.}
    \label{Matrix}%
    \end{figure*}

To build insight into the structure of synthetic images with a rotating reflecting surface, we first look at an analytic model of a Keplerian torus. The model is described in Appendix A of \citep{2019A&A...632A...2D}. We use a single torus in the equatorial plane with radius $r_{\rm ring} = 5$ $GM/c^2$ and cross-sectional thickness of $1$ $GM/c^2$. The model is perfectly optically thin (so absorption is not taken into account) and we use as emissivity profile a step-function that is $1$ inside the torus and $0$ outside it. Additionally, a Keplerian velocity vector is given to the material, to take the relativistic boosting effect into account. Synthetic images for different surface radii $R$ are shown in Fig. \ref{Rad_radius} and for different spins $a$ in Fig. \ref{Rad_spin}. In Fig. \ref{Matrix}, synthetic images with several spins $a$ and $i$ are shown, which will be used in the GRMHD models as well.

First of all, one can see the lensed image of the torus and subsequent higher order lensed images interior to that converging to the photon sphere. The latter corresponds to geodesics orbiting the surface increasingly many times.

Then we immediately see, that in all images, a ring in the center and several (partial) rings subsequently exterior to that can be seen. These are reflected images of the emitting torus in the rotating reflecting surface. The central ring corresponds to the direct reflection, with which we mean that the geodesic has a deflection angle in the $\theta$-direction around $0$. The subsequent rings outside that, correspond to geodesics with deflection angles in the $\theta$-direction of half-integer multiples of an orbit larger. This is expected from the results of section \ref{Light_paths}. Around the middle, there should be a geodesic with deflection angle $0$ and when going outwards toward the photon ring, the deflection angle will increase or decrease depending on the side.

The number of rings depends on the surface radius $R$ and spin $a$, since those decide whether or not the surface radius $R$ lies (partially) outside the photon sphere or not. In Fig. \ref{Rad_radius}, one can see in the first two plots, that not all rings are full. In the first plot, all rings except the central one are open in the top-left. In the second plot, the first ring exterior to the central ring has become full, but the next rings are open in the top-right. This is because the surface lies outside the photon sphere there and limits the possible angular distance geodesics can travel, thereby blocking higher-order rings. In the later plots, the surfaces sink below the photon sphere and an infinite amount of rings are created converging to the photon sphere. Similar behavior is possible by increasing the spin, where part of the surface can start lying above the photon sphere from a certain spin and higher.

In Fig. \ref{Rad_radius}, we see that the size of the rings decreases when the radius decreases. The rings seem to converge in the later plots. This is expected, since geodesics with low deflection angles converge fast with decreasing radius.

In Fig. \ref{Rad_spin}, we see that the sizes of the rings decrease and that their shape change. The first is caused by the fact that the radius decreases with higher spin. The second is because of the frame-dragging. On the left, the rings are close to the center, caused by geodesics moving along with the rotation. On the right, the rings are farther from the center, caused by geodesics moving against the rotation. The rings seem to converge in the later plots. This is expected again, since geodesics with low deflection angles converge fast with increasing spin.

In Fig. \ref{Matrix}, images are shown with a negative spin, meaning that the rotation of the surface is the opposite of that of the torus. Here, the structure of the rings is mirrored from that of positive spins.

Also, different inclinations are shown. It can be seen that for low inclinations, the rings are almost concentric circles, which is expected from the symmetry in the spin axis.

\subsection{General Relativistic Magnetohydrodynamical Model}
\begin{figure*}
   \centering
   \includegraphics[width=\textwidth]{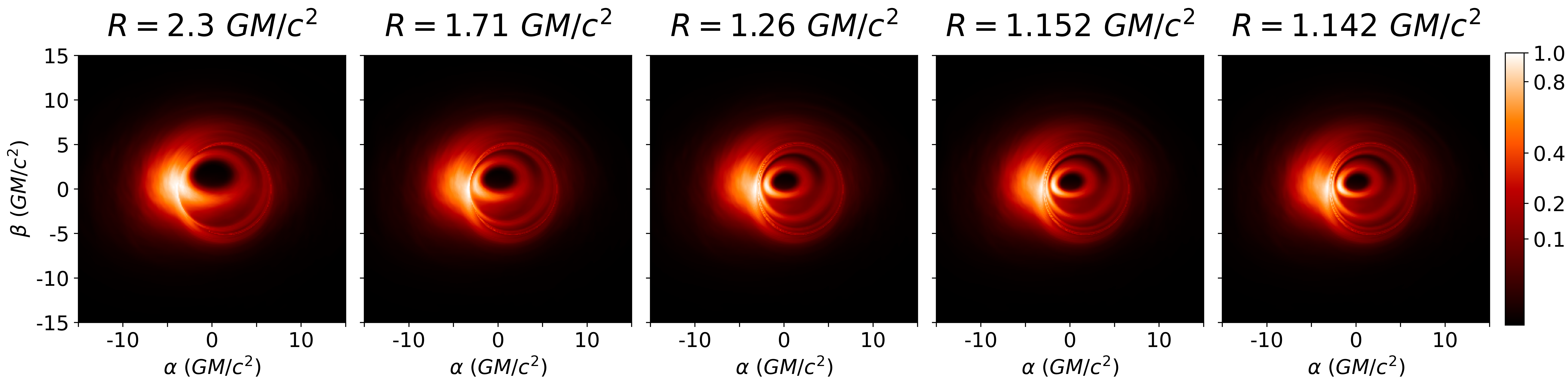}
   \caption{Synthetic images of rotating reflecting surfaces with different radii $R$, with spin $a = 0.99$, inclination $i = 60 ^{\circ}$ and using a SANE GRMHD model. The radii correspond from left to right to $2.0$, $1.5$, $1.1$, $1.01$ and $1.001$ times the radius $r_+$ an event horizon would have had.}
    \label{GRMHD_Radius}%
    \end{figure*}
\begin{figure*}
   \centering
   \includegraphics[width=\textwidth]{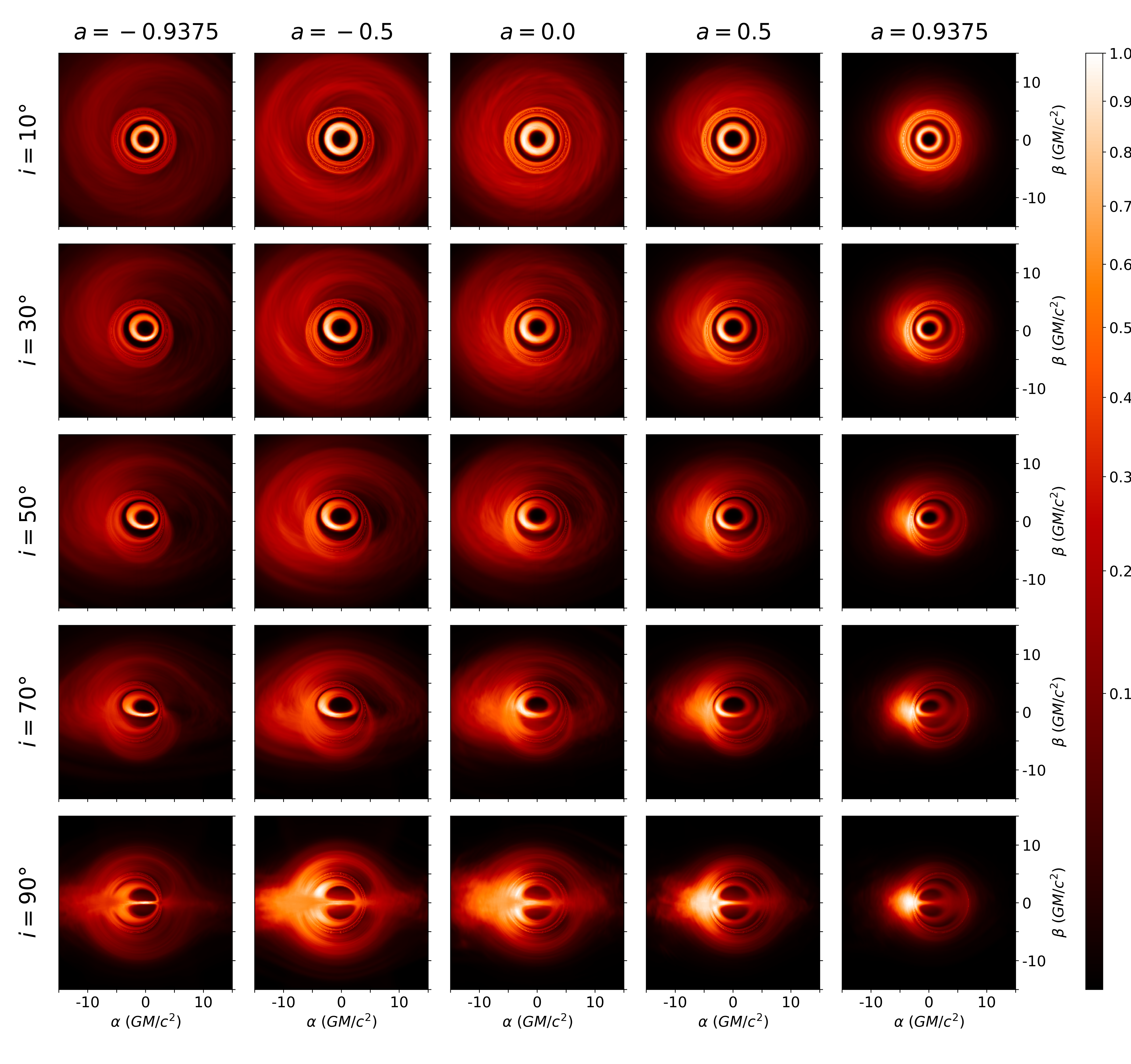}
   \caption{Synthetic images of rotating reflecting surfaces with different spin a and inclination i, using a SANE GRMHD model. The surface radius is $R = 1.01 r_+$, where $r_+$ is the radius an event horizon with the corresponding spin would have had.}
    \label{GRMHD_matrix}%
    \end{figure*} 
\begin{figure}
   \centering
   \includegraphics[width=9cm]{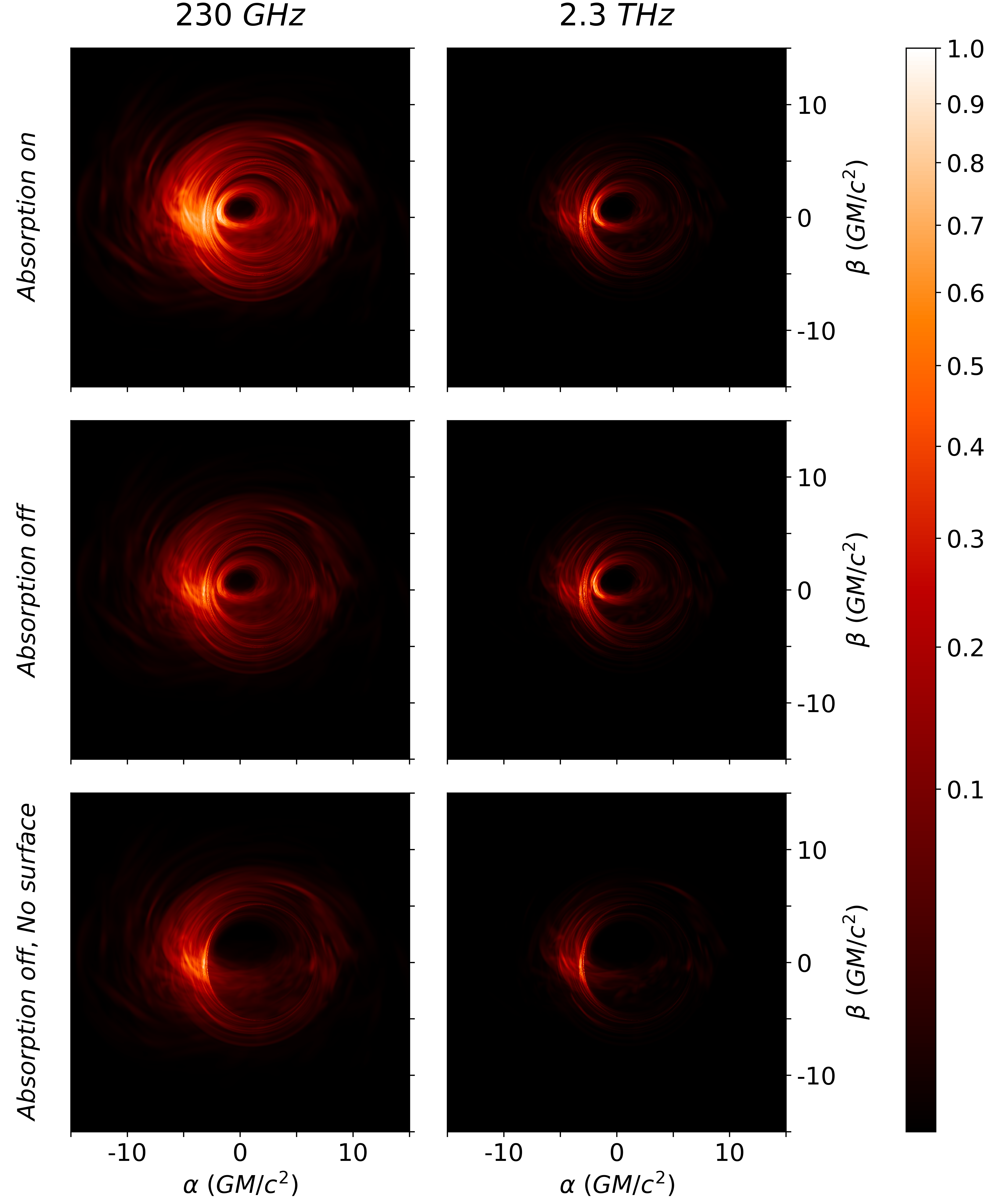}
   \caption{Synthetic images using one snapshot of a SANE GRMHD model of a rotating reflecting surface with the absorption on and off and a black hole with the absorption off for two frequencies. The spin is $a = 0.9375$. The radius of the surface is $R = 1.01 r_+$, where $r_+$ is the radius an event horizon with the corresponding spin would have had. The inclination is $i = 60 ^{\circ}$.}
    \label{Snapshots}%
    \end{figure}
Now that we understand the structure of synthetic images of rotating reflecting surfaces, we move on to a more realistic model: a GRMHD simulation. For this, SANE 3D-GRMHD simulations by the GRMHD code {\tt BHAC} \citep{Porth_2017, 2019A&A...629A..61O} were used that were made for the EHT GRMHD simulation library \citep{2019ApJ...875L...1E}. The simulations have spins $a \in \{0,\ \pm0.5,\ \pm0.9375\}$. As initial condition a Fishbone-Moncrief torus \citep{1976ApJ...207..962F} is used of which the rotation axis is aligned with the spin axis of the black hole. A numerical grid with three spatial coordinates $r, \theta$ and $\phi$ is used in spherical modified Kerr-Schild coordinates. The simulation domain ranges from the horizon to a radius of $3333\ G M/c^2$. The spatial resolution of the grid is $N_r \times N_{\theta} \times N_{\phi} = 512 \times 192 \times 192$.

Important to note is that these simulations are considered black holes. As described earlier, the surface is 'added' in the radiative transfer code. Since the calculated geodesics are not allowed inside of the surface, the region inside the surface radius $R$ is essentially cut out. Because the GRMHD simulations do not contain a surface, the effects of deceleration of material falling onto the surface are not taken into account. Emission related to heating
of the surface due to the release of kinetic energy by the infalling matter has also been neglected, in line with a scenario of a surface with a very large heat capacity as it is assumed for certain black hole mimickers \citep{Mazur2004}.

The emission model that is used here is that of synchrotron emission and self-absorption based on the properties of the plasma as calculated in the GRMHD simulation. The images are integrated over a simulation time from $t = 9500 - 10000$ $M$ by taking the average over 51 snapshots separated $10$ $M$ in time each. The electron temperature description from \citep{2016A&A...586A..38M} is used with $R_{\rm high} = 3$. This is chosen to reflect the analytic Keplerian torus model and to guarantee good sight at the shadow region. Reflecting features are expected for large values of $R_{\rm high}$ as well, so when the emission is jet-dominated, but may be more obscured by the jet. We use the mass and distance to Sgr A* and fit the accretion rate by demanding an integrated flux of $2.4$ Jy at 230 GHz. In Fig. \ref{GRMHD_Radius}, synthetic images are shown for different surface radii $R$. In Fig. \ref{GRMHD_matrix}, synthetic images with several spins $a$ and $i$ are shown. Fig. \ref{GRMHD_Radius} can be compared to Fig. \ref{Rad_radius} and Fig. \ref{Matrix} with Fig. \ref{GRMHD_matrix}.

Again, we clearly see the accretion disk, a central ring and several rings exterior to that. The sizes, shapes and positions are mostly the same as those for the analytic Keplerian torus model. The central ring in the images of the GRMHD model seems to be a bit thicker. This is caused by the geometrically thicker accretion disk.

New in the images from the GRMHD model is the clear flux distribution. Surprisingly, the central ring is now the brightest object in the image. This is caused by Doppler-boosting of the gas rushing in radially towards the surface. Normally, this high intensity light is not seen because it ends up falling in the horizon. In this case, however, the light is reflected and able to escape. This radial boosting effect can be shown to be the explanation in three ways. Firstly, the brightening of the ring cannot be seen in the analytic torus model, which has as main difference with the GRMHD model that the gas has no radial velocity. Secondly, from \citep{2019ApJ...885L..33N} it is known that inward radial velocity enhances the depth of the shadow of a black hole. When you include a reflecting surface, the post-reflecting segment of the geodesic (when tracing backwards in time) sees radiating gas moving rapidly "towards" the observer, so we expect a huge boost. Thirdly, when separating the contributions to the intensity from the pre- and post-reflection parts of the geodesic, you clearly see the intensity is dominated by the post-reflection intensity.

In most images, the central ring is brighter on the left side. This is caused by Doppler boosting just as in the accretion disk. The exception to this behavior are the images with spin $a = -0.9375$. Here the ring is the brightest on the bottom. We speculate that this behaviour is caused by the retrograde motion of the accretion disk canceling the boosting effect of spacetime to a large degree.

In addition to the central ring, again higher order rings exterior to the central one can be distinguished.

All these images were averaged to clearly show the reflecting features. However, accretion near supermassive black holes is highly variable. It is therefore interesting to look at an image using only a single snapshot as well. This is shown in Fig. \ref{Snapshots} for a snapshot at time $10000$ M and with a resolution of $1000 \times 1000$ pixels. To study the influence of frequency and optical depth, the image is shown for frequencies of respectively $230$ GHz and $2.3$ THz and with respectively the absorption on and off. For completeness, to compare the images to those of a black hole, also images with absorption off and without a surface are shown. 

Firstly, the images are less smooth than averaged images and show a lot more structure in the accretion disk and its reflections. Despite this, the central ring and some higher-order rings are clearly visible. To some extent, it is even possible to couple structures in the accretion disk with their reflected version. This would lead to interesting light echoes.

To study the optical depth, one can compare the images of the reflecting surfaces with the absorption on and off. At $230$ GHZ, one sees that the image is (somewhat) optically thick in some regions. However, it is expected that these regions will become optically thin when increasing the frequency. This is exactly what can be seen in the images at $2.3$ THZ where there is almost no difference anymore between having the absorption on and off.

Finally, when comparing the two frequencies, one can see that at a higher frequency the reflection becomes the dominate feature in the image. This trend continues for higher frequencies. This means that for a reflecting surface and black hole with the same intensity at $230$ GHz, the intensity at higher frequencies will be higher for reflecting surfaces. So reflecting surface models are a way to increase the spectrum of black hole candidates at higher frequencies. 

\section{Discussion}\label{Discussion}
\subsection{Static Spherical Surfaces}
We can compare our results on static spherical reflecting surfaces with those in \citep{2022ApJ...930L..17E}. There the reflection law given in Eq. \ref{Reflection_law_EHT} was used. This is the exact same reflection law that we used, namely Eq. \ref{Reflection_law_static}. The latter was derived from a general description for reflecting surfaces in arbitrary spacetimes. This strengthens our confidence in the correctness of the model and gives context and insight into its origin and validity.

For the synthetic images, \citep{2022ApJ...930L..17E} used a long-duration simulation of a hot accretion flow in the MAD state around a black hole with spin $a = 0$ \citep{2022MNRAS.511.3795N} averaged over a long timescale. For the electron temperature the same prescription as here was used with parameter values $R_{\rm high} = 20$ and $R_{\rm low} = 1$. Likewise, the accreting gas density was scaled such that the observed flux density at $230$ GHz is equal to $2.4$ Jy. Finally, an inclination of $i = 60 ^{\circ}$ and a reflecting surface radius of $R = 2.5$ $GM/c^2$ were used. An image with albedo $A = 1$ can be seen on the left in the second row of Fig. 15 in \citep{2022ApJ...930L..17E}.

We can compare this image to the first image in Fig. \ref{Rad_spin} and the images in the central column of Fig. \ref{Matrix} and Fig. \ref{GRMHD_matrix}. Apart from the surrounding accreting material, the images look very similar. In both images, a bright central ring and a second ring exterior to that are clearly visible. The position, shape, size and flux distribution along the rings are the same. This is all despite the use of different accretion and emission models. This gives confidence in the validity and generality of both results.

\subsection{Comparison Static and Rotating Surfaces}
We developed a model for rotating reflecting surfaces with the corresponding reflection law given in Eq. \ref{Kerr_reflection}. Surprisingly enough, this is the same as that for static surfaces: Eq. \ref{Reflection_law_static}. This result depends of course on the coordinates used; the Boyer-Lindquist coordinates are coordinates that nicely extend the reflection law from the static to the rotating case. Here one has to remind oneself that the model for the rotating surface is an approximation and the reflection law may be different for a metric and surface of a specific mass distribution.

The synthetic images showed that in the rotating case, we see again ring-like features inside the photon ring. The location, shape, size and flux distribution along the ring depend on the radius of the surface $R$, spin $a$ and inclination $i$. However, in all cases, there is a bright ring in the middle of the `shadow region' and one or more clearly visible rings exterior to that as well.

The ring features are the result of specular reflection. In the case of diffuse reflection, we could imagine the images to be totally different. At first sight, we would expect the surface to radiate in all directions, resulting in the illumination of the whole shadow region. There might be some substructure there caused by the accretion disk illuminating different parts of the surface more than others, but the ring features found here will probably be gone. If the light is reflected off some sort of atmosphere, we would expect the reflected light to be diffuse as well, but still mainly directed close to the direction of specular reflected light. In that case, we would expect the images to still have the ring features, although the rings will probably become a bit thicker. 

\subsection{Implications for the Event Horizon Telescope}
Since the image features of rotating reflecting surfaces are similar to those of static ones, the argument presented in \citep{2022ApJ...930L..17E} should still hold when allowing rotating. This means that for high albedo, reflecting surfaces and event horizons should clearly be distinguishable, even with a $15 \mu$as blur, and consequently that such reflecting surface models can be excluded using the EHT observations. Given that absorbing surfaces have been mostly excluded as well, this strengthens the conclusion that the shadow indeed indicates the existence of an event horizon.

Of course, it should be mentioned that this is still a toy model and still much can be improved to make more realistic models of exotic compact objects with rotating reflecting surfaces. We also note that these kinds of models might be interested in the context of neutron stars, which are thought to have (highly) reflecting surfaces \citep{1978Natur.271..216L}.

\section{Conclusion}\label{Conclusion}
In this paper, we looked at rotating reflecting surfaces in the context of testing the existence of event horizons using horizon-scale images of black holes. We have developed a general description of reflecting surfaces in arbitrary spacetimes and used this to define specific models for static spherical surfaces and rotating surfaces. The reflecting surfaces create an infinite set of ring-like features in synthetic images inside the photon ring. There is a bright central ring in the middle and higher order rings subsequently lie exterior to each other converging to the photon ring. The shape and size of the ring features change only slightly with the radius of the surface $R$, spin $a$ and inclination $i$. In all cases, this results in features inside the middle of the ‘shadow region’. We conclude that rotating reflecting surfaces have clear observable features and that the Event Horizon Telescope is able to observe the difference between reflecting surfaces and an event horizon. Reflecting surfaces with high reflectivities can already be excluded. Together with the exclusion of absorbing surfaces, this strengthens the conclusion that the shadow indeed indicates the existence of an event horizon.

\begin{acknowledgements}
      We thank Luciano Rezzolla for insightful comments regarding this research. We also thank Christiaan Brinkerink, Jesse Vos, Jordy Davelaar and Joris Kersten for helpful discussions. Finally, we want to thank the anonymous EHT Collaboration internal reviewer for helpful comments.\\ 
      This work is supported by the Radboud University and the ERC Synergy Grant `Blackholistic'.\\
      YM is supported by the Shanghai Municipality orientation program of Basic Research for International Scientists (Grant No.\,22JC1410600), the National Natural Science Foundation of China (Grant No.\,12273022), and the National Key R\&D Program of China (No.\,2023YFE0101200).\\
      HO is supported by the Center for Research and Development in Mathematics and Applications (CIDMA) through the Portuguese Foundation for Science and Technology (FCT - Fundação para a Ciência e a Tecnologia), references UIDB/04106/2020, UIDP/04106/2020. HO acknowledges support from the projects PTDC/FIS-AST/3041/2020, CERN/FIS-PAR/0024/2021 and 2022.04560.PTDC. HO has further been supported by the European Union’s Horizon 2020 research and innovation (RISE) programme H2020-MSCA-RISE-2017 Grant No. FunFiCO-777740 and by the European Horizon Europe staff exchange (SE) programme HORIZON-MSCA-2021-SE-01 Grant No. NewFunFiCO-10108625. During part of this project, funding for HO came from Radboud University Nijmegen through a Virtual Institute of Accretion (VIA) postdoctoral fellowship from the Netherlands Research School for Astronomy (NOVA). HO is supported by the Individual CEEC program - 5th edition funded by the Portuguese Foundation for Science and Technology (FCT).
      \end{acknowledgements}

\newpage
\bibliographystyle{aa} 
\bibliography{References} 


\end{document}